\documentclass[12pt]{iopart}

\usepackage{graphicx}
\usepackage{bm}

\begin{document}

\title[Thermal conductivity and phonon hydrodynamics in TMDs from 
       first-principles]
      {Thermal conductivity and phonon hydrodynamics in transition 
       metal dichalcogenides from first-principles}

\author{Pol Torres}
\address{Institut de Ci\`encia de Materials de Barcelona (ICMAB-CSIC),
             Campus Bellaterra, 08193 Bellaterra, Catalonia, Spain}

\author{Francesc Xavier Alvarez}
\address{Departament de F\'isica, Universitat Aut\`onoma de
             Barcelona (UAB), Campus Bellaterra,
             08193 Bellaterra, Catalonia, Spain}

\author{Xavier Cartoix\`a}
\address{Departament d'Enginyeria Electr\`onica, Universitat
             Aut\`onoma de Barcelona (UAB), Campus Bellaterra,
             08193 Bellaterra, Catalonia, Spain}

\author{Riccardo Rurali}
\address{Institut de Ci\`encia de Materials de Barcelona (ICMAB-CSIC),
             Campus Bellaterra, 08193 Bellaterra, Catalonia, Spain}
\ead{rrurali@icmab.es}

\date{\today}

\vspace{10pt}
\begin{indented}
\item[]November 2018
\end{indented}

\begin{abstract}
We carry out a systematic study of the thermal conductivity 
of four single-layer transition metal dichalcogenides, MX$_2$ 
(M = Mo, W; X = S, Se) from first-principles by solving the Boltzmann
Transport Equation (BTE). We compare three
different theoretical frameworks to solve the BTE beyond the
Relaxation Time Approximation (RTA), using the same set of interatomic
force constants computed within density functional theory (DFT), finding that the RTA severely underpredicts the thermal conductivity of MS$_2$ materials.
Calculations of the different phonon scattering relaxation times
of the main collision mechanisms and their corresponding mean free
paths (MFP) allow evaluating the expected hydrodynamic behaviour
in the heat transport of such monolayers. These calculations indicate
that despite of their low thermal conductivity, the present TMDs can
exhibit large hydrodynamic effects, being comparable to those of
graphene, especially for WSe$_2$ at high temperatures.
\end{abstract}

%
\vspace{2pc}

%
%
%

\section{Introduction}
\label{sec:intro}

Since the isolation of monolayer graphene~\cite{NovoselovScience04},
research in 2D materials has experienced a significant 
increase~\cite{NovoselovPNAS05}. Transition metal dichalcogenides 
(TMDs)~\cite{WangNatureNano12,ManzeliNatRev17} have attracted a huge 
interest due to their particular physical properties, such as tunable 
thickness-dependent bandgap~\cite{MakPRL10,ZhangNatNano14} or 
anisotropic response under tensile strain~\cite{Wang2014}.
Similarly to graphene, TMDs have covalent in-layer bonds and weaker 
van der Waals out-of-layer interactions and thus, beside being synthesized 
bottom-up, they can also be exfoliated, allowing the exploration of 
atomically thin layers~\cite{MakPRL10}. They have a significant
bandgap, which allows bypassing the most important limitation of 
graphene when it comes to the design and engineering of electronic 
devices~\cite{RadisavljevicNatNano11}. Their bandgap is indirect
and ranges from 0.85 to 1.23~eV in the bulk, while it increases
up to 1.5-1.8~eV and it becomes direct for the monolayer, a 
fact that paves the way for their use in photonic 
applications~\cite{SplendianiNL10,LiuAPL14}. Additionally, they are 
good candidates for applications in catalysis, energy storage, and 
sensing, at variance with graphene, which is chemically inert
unless functionalized with specific molecules~\cite{ChhowallaNatChem13}.

The thermal properties of TMDs have been comparatively less studied,
but they are attracting a growing interest~\cite{Wang2017}.
A few theoretical and experimental 
works have started to determine the properties of TMDs with high 
accuracy. The phonon dispersions were 
the first features to be studied from a theoretical 
viewpoint~\cite{Molina2011,Sahin2013,Terrones2014} and the Raman active modes 
were also experimentally reported~\cite{Terrones2014}. Further theoretical calculations 
have been able to determine the thermal conductivity from molecular 
dynamics~\cite{Liu2013,Wang2016,Norouzzadeh2017} and from lattice 
dynamics using density functional theory (DFT) through the Boltzmann 
transport equation (BTE) under the relaxation time approximation 
(RTA)~\cite{Gu2014}. Recently, more accurate solutions of the 
BTE~\cite{ShengBTE_2014} have allowed to compute with greater 
accuracy the thermal conductivity of MoS$_2$~\cite{Cepellotti2015,
Gandi2016}.

In this work we present a unified, {\it ab initio} description 
of the thermal conductivity, $\kappa$, of single-layer transition metal dichalcogenides, MX$_2$ 
(M = Mo, W; X = S, Se) by obtaining the interatomic force
constants from DFT calculations and then solving numerically the BTE 
beyond the RTA. We show that the latter can result in underprediction 
of $\kappa$ up to the 50\%. Additionally, we demonstrate that momentum conserving phonon-phonon 
scattering, {\it i.e.} normal ({\it N}) processes, are the predominant 
three-phonon collision mechanisms. This can lead to phonon 
hydrodynamics~\cite{Guyer1966a,Guyer1966,AlvarezJAP09}, a fluid-like heat transport regime 
characterized by the appearance of phenomena like Poiseuille flow and 
second sound, and whose relevance in graphene and in other 2D materials has already 
been highlighted~\cite{Cepellotti2015,Lee2015}.

\section{Computational methods}
\label{sec:methods}

Heat transport in dielectric materials can be described by the phonon BTE
\begin{equation}\label{eq_BTE}
\left(\frac{\partial n_{\mathbf{q}}}{\partial t}\right)_{\textrm{drift}} \equiv  \frac{\partial n_{\mathbf{q}}}{\partial t}+\mathbf{v}_{\mathbf{q}}
 \cdot \frac{\partial n_{\mathbf{q}}}{\partial \mathbf{r}} = \left(\frac{\partial n_{\mathbf{q}}}{\partial t}\right)_{\textrm{scattering}} \; ,
\end{equation}
which states that a perturbed distribution of phonons $n_{\mathbf{q}}$ restores its equilibrium state $n_{\mathbf{q}}^0$, expressed by the Bose-Einstein distribution function, through scattering (collision) processes~\cite{FugalloPS18}. 

This complex equation can be solved through different methods. A 
common approximation was the RTA, due to its simplicity. This model has been particularly useful as long as the size of the samples was quite big and the experiments were performed under slow heating conditions.
The improvement of manufacturing techniques and new electronic devices have allowed to reduce the sample size down to few nanometers and the heating times can be of the order of picoseconds. In these cases it has been proved that the RTA solution is often far from the real solution, leading to the proposal of more accurate solutions to the linearized BTE. In this work three of them are considered: the direct solution derived by Chaput~\cite{Chaput2013} (L-BTE), the iterative solution (I-BTE) proposed by Ward~\textit{et.al.}~\cite{Ward2009}, and the kinetic collective model (KCM) approach due to De Tom\'as~\textit{et.al.}~\cite{DeTomas2014,Torres2016}. 

The thermal conductivity derived from the L-BTE solution can be expressed in its final form as:
\begin{equation}\label{eq_LBTE}
\kappa^{\alpha \beta} (\omega) = \int \frac{\mathbf{\rho}_{\alpha \beta}(\omega ')}{\omega '- i\omega}\textrm{d}\omega ' \; ,
\end{equation} 
where $\mathbf{\rho}_{\alpha \beta}(\omega ')$ is a spectral density. For more details of the derivation of this solution we refer to the original article~\cite{Chaput2013}.

In the case of the I-BTE approach, an iterative process is done starting from the RTA solution in order to converge to the final solution. The thermal conductivity is then expressed as:
\begin{equation}\label{eq_iterative}
\kappa_{ij}=\frac{1}{k_b T^2 \Omega N}\sum_{\textbf{q}}n_{\textbf{q}}(n_{\textbf{q}}+1)(\hbar \omega_{\textbf{q}})^2 v_{i,\textbf{q}}\tau_{\textbf{q}}(v_{j,{\textbf{q}}}+\Delta_{j,\textbf{q}}) \;,
\end{equation}
where $\Omega$ is the volume of the unit cell, $N$ the number of $\textbf{q}$-points used in the sampling, $n_{\mathbf{q}}$ the Bose-Einstein distribution function, $\omega_{\textbf{q}}$, $v_{\textbf{q}}$ and $\tau_{\textbf{q}}$ correspond to the phonon frequency, velocity and relaxation time respectively. The term $\Delta_{j,\textbf{q}}$ captures the deviation on the heat current with respect to the RTA approach.
In highly kinetic materials like alloys, this difference between both approaches is close to zero, while it can be very significant in the materials we are studying here.

In order to evaluate more precisely how important this kinetic contribution is with respect to the real solution, the KCM was recently proposed~\cite{DeTomas2014,Torres2016}.
Here the total contribution to the thermal conductivity is split into a kinetic and a collective term, where the former corresponds to the effect of individual phonon collisions and the latter to a global contribution due to the coupling of phonon modes as a result of \textit{N} processes. This allows to evaluate the collective contribution to $\kappa$ in materials with relevant momentum conserving scattering processes:
\begin{equation}\label{eq_sigma}
\kappa_{ij}=\kappa^K_{ij}(1-\Sigma) + \kappa^C_{ij} \Sigma \; ,
\end{equation} 
where $\kappa^K_{ij}$ corresponds to the kinetic contribution to the thermal conductivity and $\kappa^C_{ij}$ to the collective one. The $\Sigma$ factor is a weighting term that determines the predominance of each term depending on the dominance of \textit{N} versus resistive (\textit{R}) phonon-phonon scattering processes $\Sigma =\Gamma_N/(\Gamma_N + \Gamma_R)$. The resistive collisions are those that contribute directly to the thermal resistance, such as Umklapp (\textit{U}) and impurity/mass defect scattering. The interested reader can find expressions for $\kappa^K_{ij}$ and $\kappa^C_{ij}$ in Ref.~\cite{Torres2016}.


All the previous discussed solutions of the BTE are currently implemented in open-source programs. The L-BTE can be solved by using the Phono3py package~\cite{Togo}, the I-BTE is implemented in the ShengBTE software~\cite{ShengBTE_2014} and the KCM in the {\sc kcm.py} code~\cite{Torres2017}. 

In order to determine the thermal conductivity of the present TMDs it is necessary to calculate their harmonic and anharmonic properties. To this end we have used DFT calculations as implemented in the VASP code~\cite{Kresse1996a} under the Local Density Approximation (LDA) by using projector augmented waves~\cite{Blochl1994,Kresse1999}. For the harmonic and anharmonic properties, a $6 \times 6 \times 1$ supercell with a $4 \times 4 \times 1$ \textbf{$k$}-point sampling was used. The second and third order interatomic force constants (IFC) have been calculated by the finite displacement method. The ShengBTE and Phono3py codes use the harmonic IFCs calculated from Phonopy~\cite{Phonopy}. Here interactions between all the atoms of the supercell are taken into account. To generate the displacements for the anharmonic IFCs, ShengBTE requires and external program called {\sc thirdorder.py}, while Phono3py has its own script inside the code. Regarding {\sc kcm.py}, it uses the output of Phono3py in order to calculate the thermal conductivity under its approach.

The details of the number of neighbors considered in the 3-phonon scattering processes and the \textbf{q}-point sampling used to calculate the anharmonic properties and the thermal conductivity will be discussed in the next section.


\section{Results and discussion}
\label{sec:results}

\subsection{Convergence study}
\label{sub:conv}

As an initial step of our analysis, we have conducted a detailed 
convergence study of the IFCs with respect to a few important 
parameters. We have studied (i)~the supercell (SC) size used for the 
calculation, (ii)~the number of neighbors considered in the 
anharmonic processes, (iii)~the number of {\bf q}-points
used to sample the Brillouin zone, and (iv)~the Gaussian smearing 
factor. Carefully converging all 
these parameters, as we discuss below, is essential to obtain
a reliable estimate of the thermal conductivity and to compare
the predictions of the different approaches to solve the BTE.

\subsubsection{Supercell size}
\label{sub:sc}

\begin{figure}[h!]
\includegraphics[width=0.5\columnwidth]{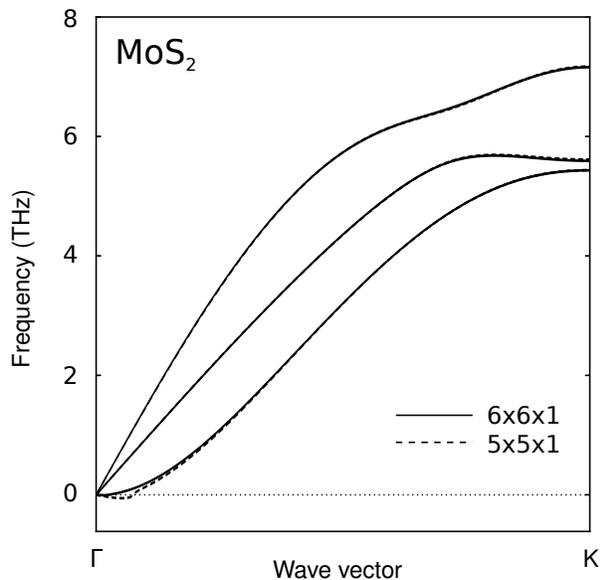}
\caption{Acoustic phonon bands for MoS$_2$ along the $\Gamma-K$ 
         direction using a $5 \times 5 \times 1$ and a $6 \times 
         6 \times 1$ supercell.}
\label{fig:SC}
\end{figure}

Two dimensional materials such as graphene or the single-layer 
TMDs here studied feature a flexural acoustic phonon band
that has a quadratic dispersion ({\it i.e.} zero sound velocity)
close to $\Gamma$. Preliminary calculations
showed that it is quite challenging to describe well the curvature
of this phonon band as it approaches $\Gamma$ in TMDs, and that the results 
critically depend on the SC size. Therefore, we took this
as a criterion to establish which is the minimum cell size
required to have well converged harmonic IFCs. Anharmonic IFCs
typically converge faster than harmonic ones and it is common
to use slightly smaller SCs for the latter~\cite{ShengBTE_2014, Raya-MorenoAPL17}. Indeed, we observe 
that providing a satisfactory description of the flexural phonon 
band turned out to be a rather stringent requirement that can be 
fulfilled only with relatively large cells (see Fig.~\ref{fig:SC}).
As a result of this convergence test we decided to use a $6 \times 
6 \times 1$ SC.

\subsubsection{Cutoff for three-phonon processes}
\label{sub:neigh} 

We used a $6 \times 6 \times 1$ SC also for the calculation of the
anharmonic IFCs by finite displacements. Computing them requires,
in principle, to carry
out a few hundreds of DFT calculations for each material.
To reduce the number of displacements considered, and given the
short range of the non-dipole contribution to the
anharmonic IFCs~\cite{Gonze1997, Wang2010},
it is sometimes
possible to calculate three-phonon collision processes
neglecting interactions between atoms beyond a given cutoff. 
Small cutoffs result in less atom triplets to study and thus
a lighter computational load, but also a lower accuracy.
Therefore, we have studied the convergence of the thermal conductivity 
of MoS$_2$ considering interactions only up to $n$-th neighbors and
increasing $n$.

For this convergence study we settled on the L-BTE solution 
implemented in Phono3py because the results should not depend on
how the BTE is solved, but rather on the range of the anharmonic 
IFCs within the DFT computational framework adopted. Additionally,
Phono3py allows generating the complete sets of displacements,
up to all neighbors, while the cutoff can later be independently specified 
when solving the BTE. Conversely, this is not possible with
{\sc thirdorder.py} as it is not straightforward to extract
from the set of displacements generated for a large number of neighbors
the set for a smaller number, as the indexing of the displacements 
changes. 

\begin{figure}[h!]
\includegraphics{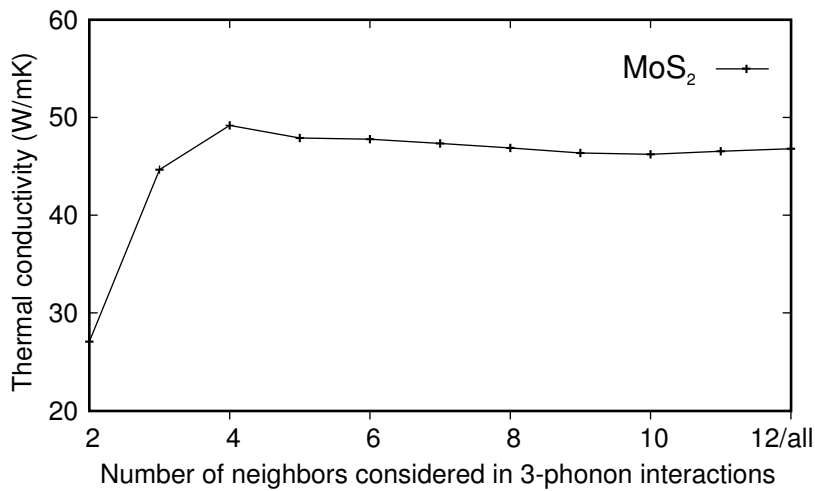}
\caption{Thermal conductivity convergence with an increasing number 
         of neighbors considered for three-phonon scattering processes 
         in MoS$_2$, using the full solution of the L-BTE from 
         Phono3py~\cite{Togo}.}
\label{fig:neigh}
\end{figure}

Our results are summarized in Fig.~\ref{fig:neigh}, where we plot 
the thermal conductivity of MoS$_2$ as a function of the number 
of neighbors. For these calculations we have used a $60 \times 60 
\times 1$ \textbf{q}-point mesh sampling, which, as shown below,
is well converged. For the $6 \times 6 \times 1$ SC adopted, the 
maximum number of neighbors that can be considered is 12. Thus,
this last value corresponds to a calculation without any limitation 
on the number of atoms in the three-phonon scattering processes,
{\it i.e.} no cutoff. It can be seen that after 5 neighbors the 
variations on the thermal conductivity are very small. We therefore
considered interactions up to 6$^{\rm th}$ neighbors for the 
calculation of the anharmonic IFCs.

\begin{figure}[h!]
\includegraphics[width=\columnwidth]{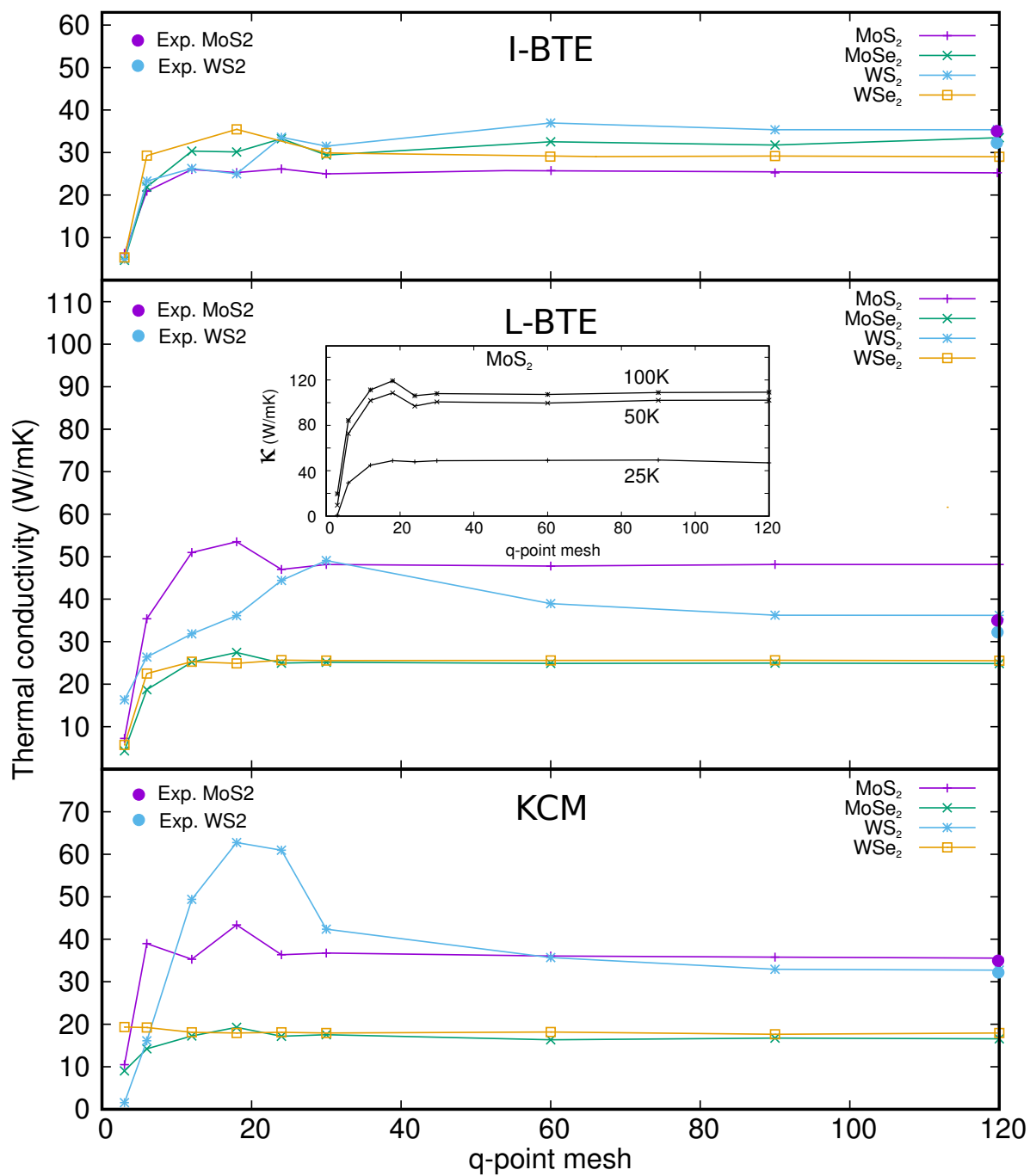}
\caption{Thermal conductivity convergence with respect to the 
         {\bf q}-point mesh for the I-LBTE, L-BTE and KCM.}
\label{fig:qpoints}
\end{figure}

\subsubsection{Brillouin zone sampling}
\label{sub:qpoints}

Once the harmonic and anharmonic IFCs are properly calculated,
following the convergence prescriptions described in Sections~\ref{sub:sc}
and \ref{sub:neigh}, the BTE must be solved on a grid of {\bf q}-points
that samples the Brillouin zone. Therefore, the next step of our
preliminary convergence study is to check the dependence of the
thermal conductivity with respect to the number $N_{\bf q}$ of 
{\bf q}-points in an $N_{\bf q} \times N_{\bf q} \times 1$ grid. We 
perform this test at room temperature, and our results are 
shown in Fig.~\ref{fig:qpoints}. Note that in this case, as we are 
dealing with the way the BTE is solved, we carried out the convergence 
test with the three approaches considered, {\it i.e.} L-BTE, 
I-BTE, and KCM. As seen in Fig.~\ref{fig:qpoints}, a $60 \times 60 \times 1$ mesh provides
converged results in all cases, and this is the mesh we took
for all systems and models. Here we have limited the phonon
mean free path (MFP) to 1~$\mu$m in order to work with a sample size
of the order of the available experimental data~\cite{Yan2014, Peimyoo2015},
reducing at the same time the computational cost. 
The value of $N_{\bf q}$ needed to converge the thermal
conductivity depends on the temperature and at lower temperatures,
where long wavelength phonons are important for heat transport,
finer {\bf q}-point grids might in principle be required. In the inset
of Fig.~\ref{fig:qpoints} we plot $\kappa$ of MoS$_2$ calculated with L-BTE as a function of
$N_{\bf q}$ at 25, 50 and 100 K. It can be observed that the limitation of
1~$\mu$m also ensures convergence with a $60 \times 60 \times 1$ grid at 
these lower temperatures as well. Notice, however, that these conclusions should
be revised for boundary scattering conditions corresponding
to larger samples, where finer {\bf q}-meshes might be needed,
particularly at low temperatures.

\subsubsection{Gaussian smearing factor}
\label{sub:gauss}

Another tunable parameter that governs the thermal conductivity is 
the smearing factor used to define the Gaussian function that modulates 
energy and momentum conservation in the three-phonon scattering processes. 
For the I-LBTE solution it has been checked that the iterative process 
only converges in all cases for $\sigma \leq 1$~cm$^{-1}$ for {\bf q}-point 
meshes up to $60 \times 60 \times 1$. The need to reduce $\sigma$ to
converge the solution of the I-BTE was previously reported in the study
of high-pressure polymorphs of silica~\cite{AramberriPRB17}. In the 
case of the L-BTE, the solution is not that sensitive and it converges 
up to $\sigma \leq 10$~cm$^{-1}$. Therefore we have used these two 
values for each model. In the case of the KCM, its solution will 
converge in all cases and thus we have used the same values as in 
the L-BTE solution, as {\sc kcm.py} uses its output.

\subsection{Thermal conductivity}
\label{sec:kappa}

\begin{figure}[t]
\includegraphics[width=1\textwidth]{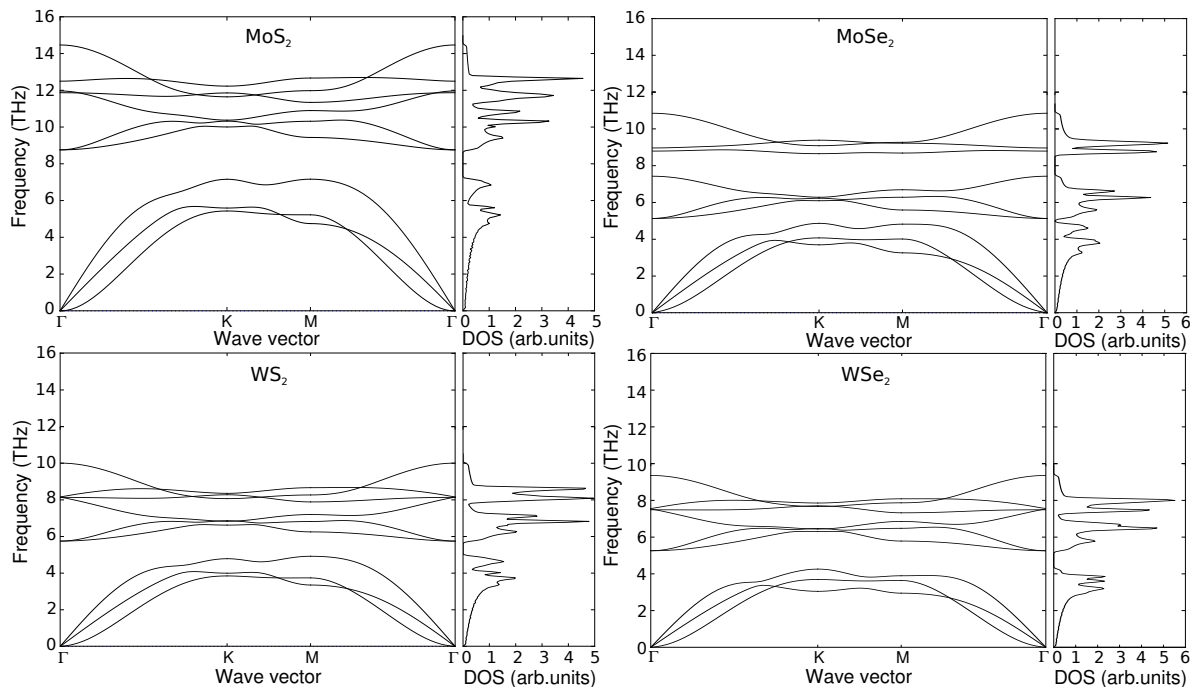}
\caption{Phonon dispersion and density of states (DOS) of MoS$_2$, 
         MoSe$_2$, WS$_2$ and WSe$_2$.
         }
\label{fig:bands}
\end{figure}

The harmonic properties of the four TMDs investigated are shown in 
Figure~\ref{fig:bands}. Despite an overall similar shape of the 
dispersion relations there are some quantitative differences. The most 
important one concerns the sound velocities of Se-based materials,
which are lower than those of S-based materials. As discussed below this 
plays an important role in the determination of $\kappa$.
Interestingly, MoSe$_2$ exhibits some difference from the
common pattern of the other materials: it has a very small 
acoustic-optical gap and its optical phonons are split into two 
subbands. However, these features do not result in a
measurable difference in the thermal conductivity, and
$\kappa(\omega)$ of MoSe$_2$ and WSe$_2$ are at a first
approximation indistinguishable (see below). 

\begin{figure}
\includegraphics[width=1\textwidth]{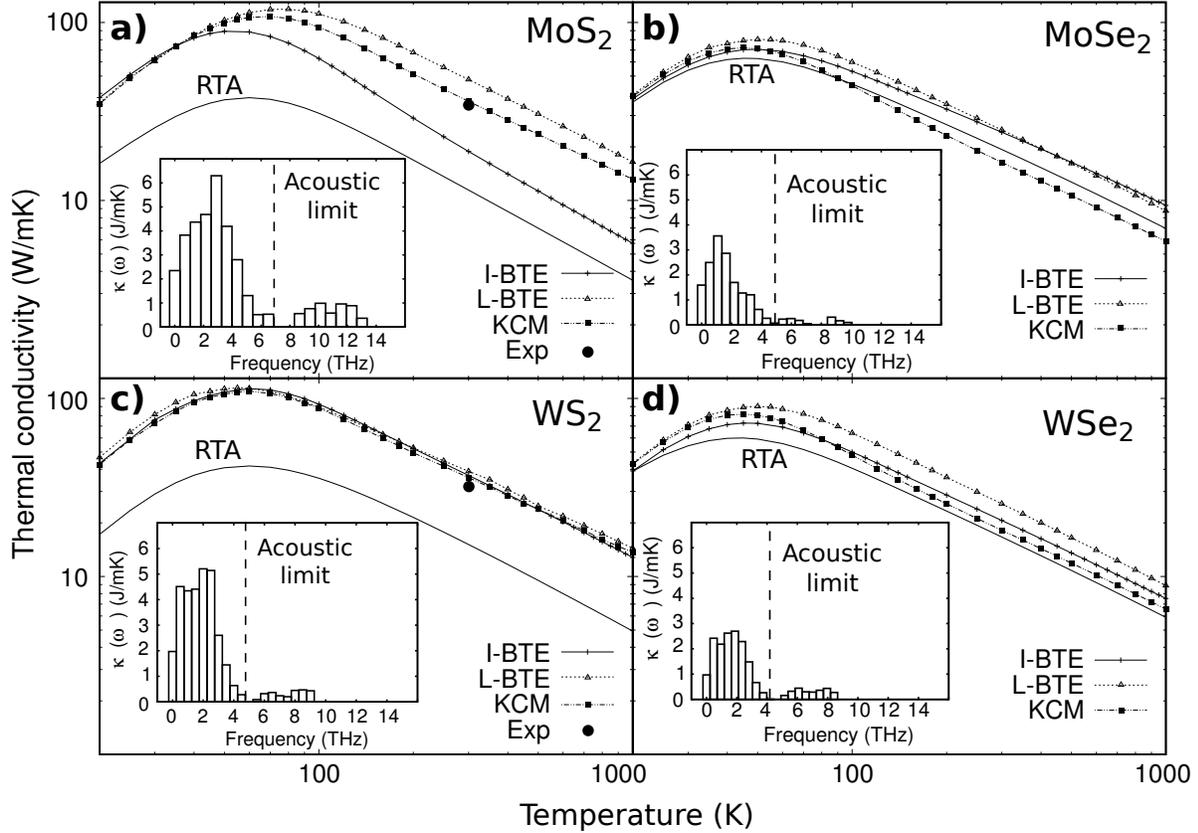}
\caption{Thermal conductivity as a function of temperature of 
         single-layer (a)~MoS$_2$, (b)~MoSe$_2$, (c)~WS$_2$ and 
         (d)~WSe$_2$ using the I-LBTE, L-BTE and KCM models. The RTA
         solution is also included for comparison. The
         experimental room-temperature values for MoS$_2$ 
         (Ref.~\cite{Yan2014}) and WS$_2$ 
         (Ref.~\cite{Peimyoo2015}) are also shown.
         The participation ratios, {\it i.e.} the frequency 
         resolved thermal conductivity $\kappa(\omega)$, at 300~K 
         using the L-BTE model are plotted in the insets.}
\label{fig:kappa}
\end{figure}

The thermal conductivities as a function of temperature are shown
in Figure~\ref{fig:kappa}. As it can be seen, the different numerical
solutions of the BTE provide estimates of $\kappa$ that can differ
from each other. As a general rule the iterative 
solution of the BTE yields lower values for $\kappa$ than those 
obtained from the L-BTE. Nevertheless, while these differences are
negligible in the case of WS$_2$ and are within 2-5~W~m$^{-1}$~K$^{-1}$  
at 300~K for MoSe$_2$ and WSe$_2$, they are very large for MoS$_2$.
The experimental room temperature value of $\kappa$ for WS$_2$ is 
well predicted by all methods, while the value for MoS$_2$ falls
in between the results obtained from I-BTE and L-BTE. The most
accurate prediction for MoS$_2$ is provided by the KCM approach. The underestimation 
of $\kappa$ as obtained from the iterative solution of the BTE
may depend, as previously discussed, on the use of the RTA solution
as initial guess. In materials with low atomic masses where \textit{N} scattering plays an 
important role, the RTA solution can be sufficiently far from the
exact solution and the iterative process might not converge~\cite{relaxons}.
In the present case, for instance, the RTA solution for MoS$_2$ is only 33.1~\% of the experimental value,
while for WS$_2$ is 47.2~\%. From Figure~\ref{fig:kappa} it can be also observed
that the RTA solution behaves different for S- and Se-compounds. The
underestimation for S-based materials is higher due to the lighter 
weight of the sulphur atoms (less than half of selenium), in agreement 
with previous results for diamond~\cite{Ward2009, Torres2016} and 
graphene~\cite{Cepellotti2015, Torres_book}, where the low weight 
of the carbon atoms and the importance of \textit{N} scattering 
makes the RTA approach underestimate the thermal conductivity.

\begin{figure}
\includegraphics[width=0.9\textwidth]{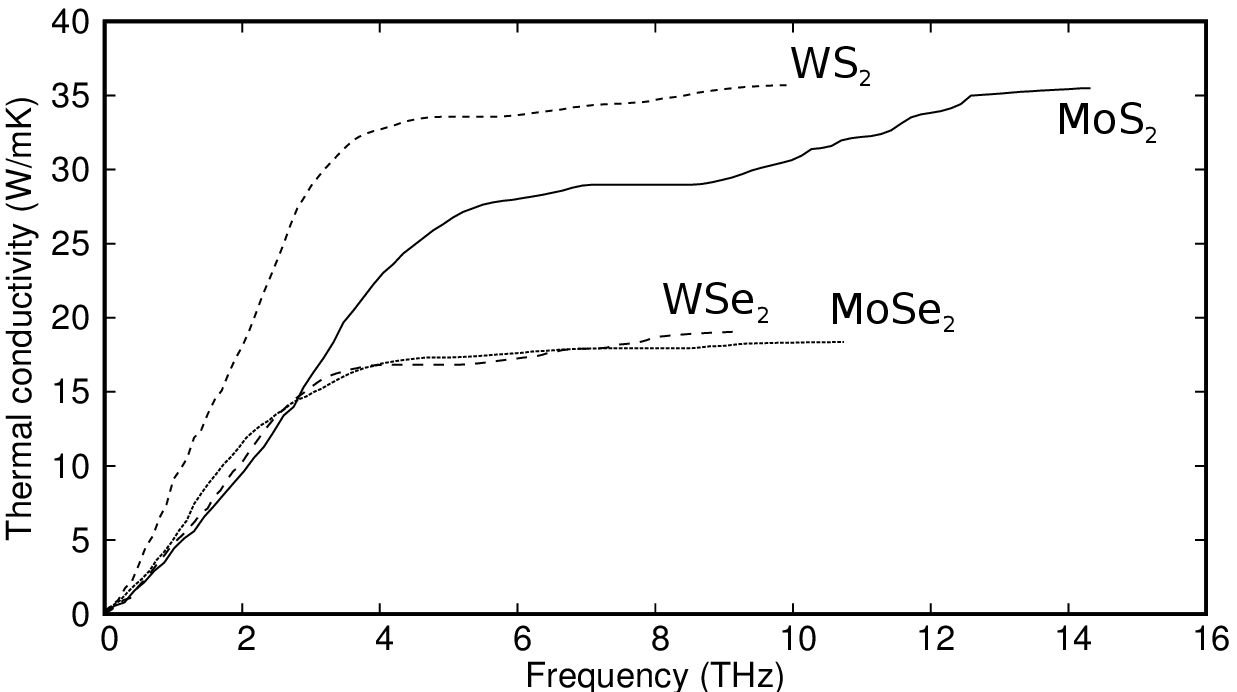}
\caption{Accumulated participation ratios $\int_0^{\omega} 
         \kappa(\omega') d\omega'$ at 300~K of MoS$_2$, MoSe$_2$, 
         WS$_2$ and WSe$_2$ using the I-LBTE model.  
         }
\label{fig:accu}
\end{figure}

To gain a better insight on which parts of the phonon spectrum
contribute more to the thermal conductivity at room temperature and to allow an easier
quantitative comparison among the different materials, we have
calculated the participation ratio, i.e. the modal decomposition
$\kappa(\omega)$, which is shown in the insets of Figure~\ref{fig:kappa}. As it
can be seen there, the thermal conductivity is almost entirely 
determined by the acoustic modes. This contribution is larger for S-based
materials, partly because of the higher phonon velocities, and thus results in
larger values of $\kappa$. The accumulation function of the participation
ratios plotted in Figure~\ref{fig:accu} clearly shows that there
are two groups of materials: S-based compounds with a $\kappa$ of
$\sim 35$~W~m$^{-1}$~K$^{-1}$ and Se-based compounds with a $\kappa$ of
$\sim 15$~W~m$^{-1}$~K$^{-1}$. MoS$_2$ is the only material where a non
negligible contribution to $\kappa$ comes from optical phonons with 
frequency between 8 and 12~THz, a feature that can also be seen in the 
corresponding inset in Figure~\ref{fig:kappa}. The similarity 
between WSe$_2$ and MoSe$_2$ is noteworthy, not only for the value of 
$\kappa$, but also for its modal decomposition.

\subsection{Phonon hydrodynamics}
\label{sec:hydro}

In thermal transport, a hydrodynamic heat transport regime is expected when momentum conservation in the phonon distribution is important. This can happen, for example, when momentum conserving collisions are dominant when compared to resistive (\textit{R}) processes~\cite{Peierls,Guyer1966a}. The first group are the so-called normal (\textit{N}) processes~\cite{Ziman}. The second group includes the intrinsic collision processes that contribute directly to the thermal resistance, i.e., Umklapp (\textit{U}) and impurity/mass defect scattering (\textit{I}). In order to determine the relaxation time associated to each scattering mechanism we take advantage of {\sc kcm.py} code~\cite{Torres2017}, which provides the averaged values of these phonon scattering relaxation times~\cite{Torres2016} as a function of temperature.

\begin{figure}
\includegraphics[width=0.7\columnwidth]{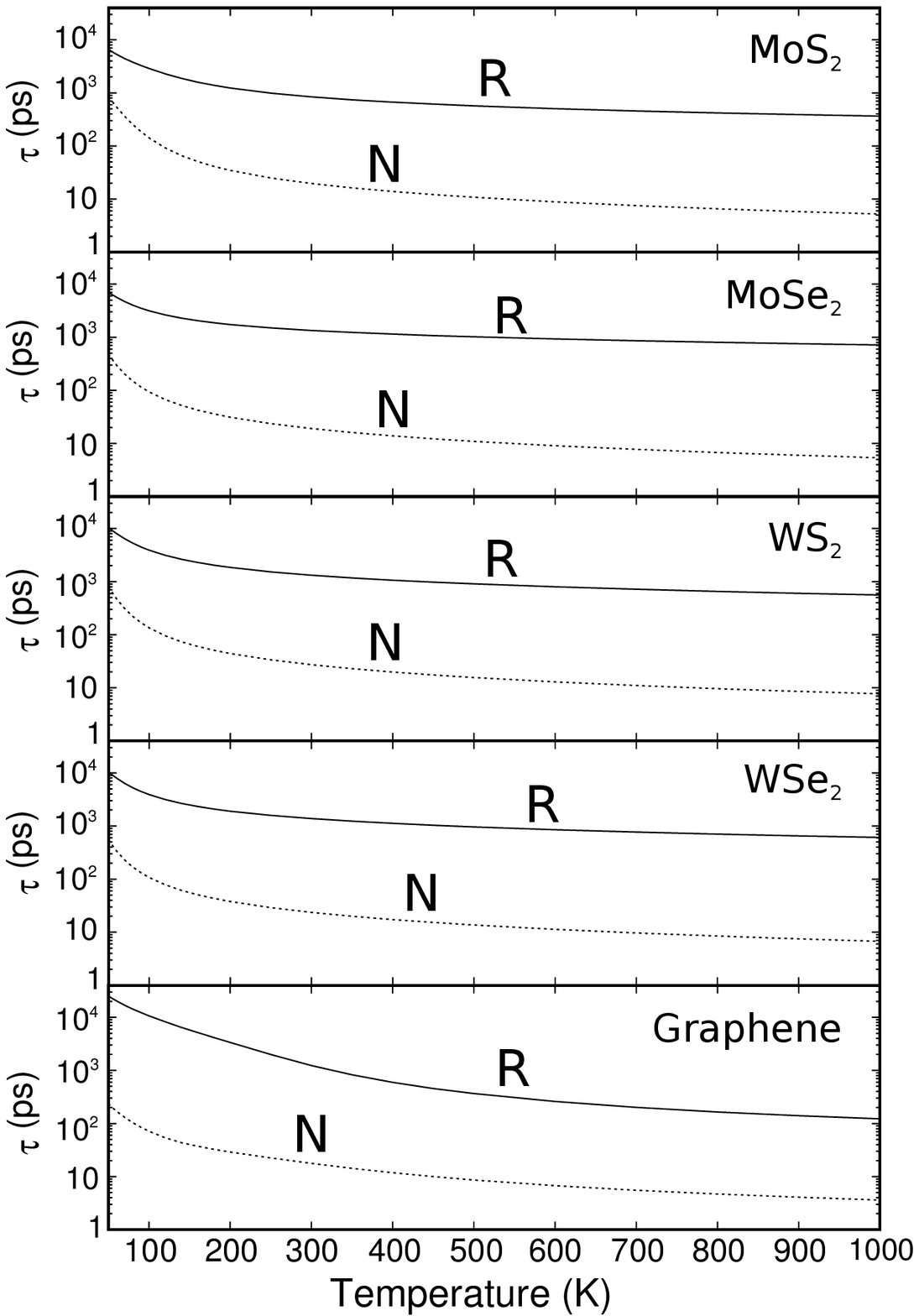}
\caption{Normal (\textit{N}) and resistive (\textit{R}) including Umklapp (\textit{U}) and 
         impurity (\textit{I}) relaxation times for MoS$_2$, MoSe$_2$, WS$_2$ 
         and WSe$_2$. Graphene is also included for 
         comparison.}
\label{fig_taus_kcm}
\end{figure}

The relaxation times for momentum conserving collisions (\textit{N}) and resistive processes (\textit{U+I}) are represented in Fig.~\ref{fig_taus_kcm}. From this representation it can be observed that for all materials the trend of both processes as function of the temperature is quite similar. While the relaxation times are almost constant at high temperatures, as the temperature decreases phonons recover their equilibrium state slower, i.e. longer relaxation times.
Here it is important to notice that in all cases there are almost two orders of magnitude of difference between the \textit{R} and \textit{N} processes, that means that \textit{N} processes dominate heat transport at all temperatures. The \textit{N} and \textit{R} relaxation times for graphene are also represented for comparison, showing a very similar trend. It can be also observed that at low temperatures the differences between \textit{N} and \textit{R} scattering for the TMDs are smaller, as the \textit{I} scattering becomes relevant. For graphene, due to the lower isotopic abundance this is not observed. These results are in agreement with previous calculations of graphene and MoS$_2$~\cite{Cepellotti2015}.

In recent articles it has been discussed that the dominance of \textit{N} collisions, as observed in Fig.~\ref{fig_taus_kcm}, can be the origin of collective effects and hydrodynamic heat transport~\cite{DeTomas2014,Cepellotti2015,Torres2016,Amir2017}. Therefore we analyze the contribution of this regime to the thermal conductivity through the splitting of the thermal conductivity proposed by the kinetic-collective approach~\cite{DeTomas2014, Torres2016}.  

\begin{figure}[h!]
\includegraphics[width=\columnwidth]{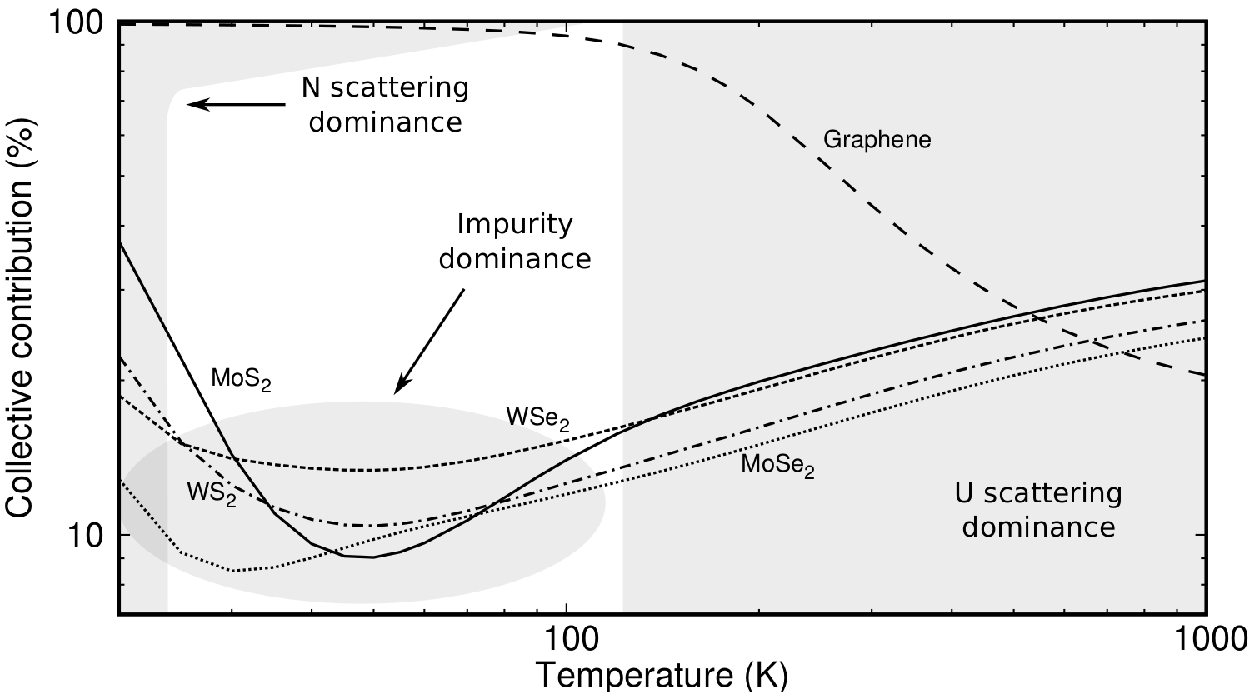}
\caption{Collective contribution to the total thermal conductivity 
         according to the KCM for the four studied single-layer TMDs 
         as well as for graphene.}
\label{fig_collective}
\end{figure}

In Fig.~\ref{fig_collective} the collective contribution to thermal transport, $\kappa^C \Sigma / \kappa $, is represented. For all materials and at all temperatures, the collective contribution to the thermal conductivity is very significant. It can be observed that all the TMDs show a very similar contribution in all the temperature range, being a 20-30~\% at high temperatures. In a range of temperatures between 25-60~K, impurity scattering corresponding to natural isotopic abundances dominates the thermal transport, producing a decrease of the collective contribution to ${\kappa}$.
This decrease is sharper in the case of MoS$_2$ and WS$_2$. As expected for this kind of 2D materials, and due to the dominance of \textit{N} processes, especially at very low temperatures, the collective contribution increases as the temperature goes down below 25~K (when impurity effects vanish), achieving values as high as 40~\% at 20~K and tending to 100~\% when T $\rightarrow 0$. The collective contribution to $\mathbf{\kappa}$ of graphene is quite different. While at high temperatures it is around 20~\%, the impurity scattering has a negligible effect allowing to \textit{N} scattering dominating the thermal transport as the temperature goes down. Notice that at room temperature such contribution is around 40~\% of the total $\mathbf{\kappa}$, and below 90~K it is more than 90~\%.

Another parameter that has been useful to explain recent experiments in a hydrodynamic framework is the so-called non-local length~\cite{Guyer1966a, Guyer1966, Amir2017, Torres2018a}. By solving the BTE [Eq.~(\ref{eq_BTE})] in a moment basis, one can obtain a hydrodynamic-like equation connecting the heat flux $\mathbf{Q}$ to the thermal conductivity $\mathbf{\kappa}$ through a non-local length $\ell$ related to the heat viscosity of a certain material~\cite{Guyer1966a, Guyer1966}:

\begin{equation}\label{eq_GK}
\mathbf{Q} - \ell^2 \left( \nabla^2 \mathbf{Q}+2 \nabla \nabla \cdot \mathbf{Q} \right) = - \mathbf{\kappa} \mathbf{\nabla} T \; .
\end{equation}

Originally this equation was derived by Guyer and Krumhansl~\cite{Guyer1966a, Guyer1966} for a pure collective regime in which \textit{N}$\gg$\textit{R}, leading thus to a collective non-local length $\ell_C$. In more recent works, this parameter has been generalized to a first order to explain experiments where \textit{N}$\sim$\textit{R}~\cite{Amir2017, Torres2018a}, leading to a kinetic contribution to the non-local length $\ell_K$.

\begin{figure}[h!]
\includegraphics[width=\columnwidth]{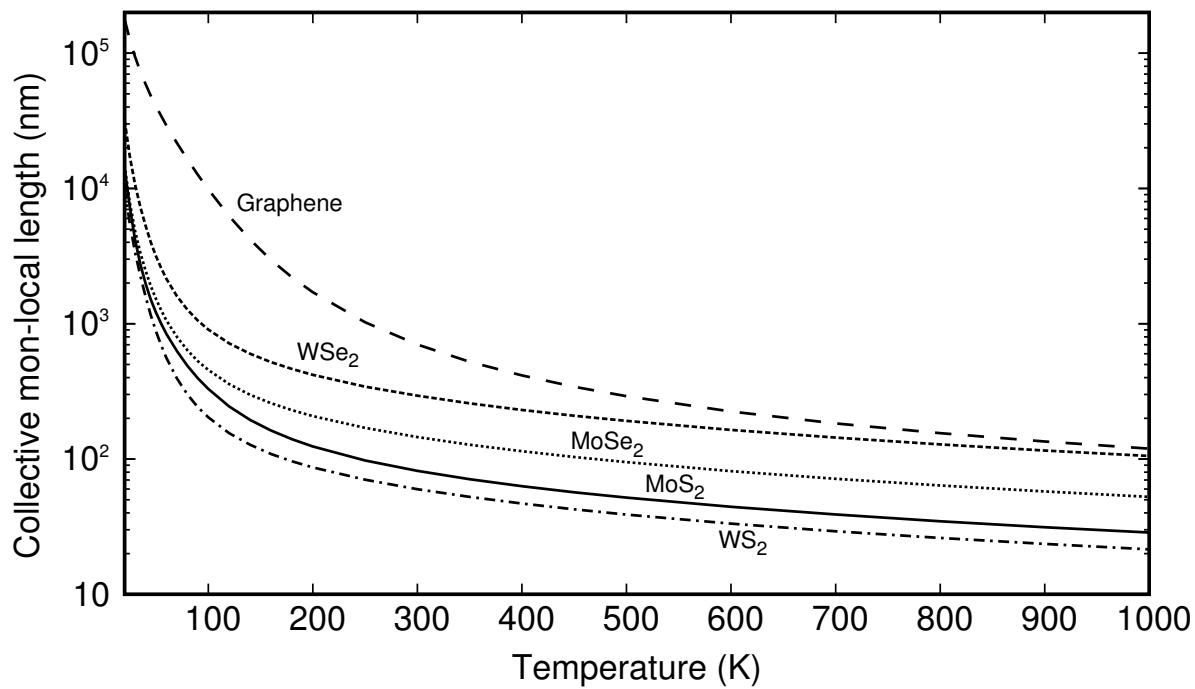}
\caption{Collective non-local length ($\ell_C$) as function of 
         temperature for MoS$_2$, MoSe$_2$, WS$_2$ and WSe$_2$. 
         The $\ell_C$ for graphene is also included for comparison.}
\label{fig_ela}
\end{figure}

The collective non-local lengths for all the studied TMDs, as well as for graphene, are represented in Fig.~\ref{fig_ela}. According to these values, hydrodynamic effects in a pure collective regime will be expected at scales between 10-100~nm at high temperatures, between 100~nm and 1~$\mu$m around room temperature and above 10~$\mu$m for T$\rightarrow 0$. Eventually, in cases where \textit{N}$\sim$\textit{R} or even when \textit{N}$<$\textit{R}, hydrodynamic effects can be observed at larger length scales due to the kinetic contribution~\cite{Amir2017, Torres2018a}. In addition, to get the full picture, it should be noticed from Eq.~(\ref{eq_GK}) that hydrodynamic effects will be observable only when both the non-local length $\ell$ and the geometric term $\left( \nabla^2 \mathbf{Q}+2 \nabla \nabla \cdot \mathbf{Q} \right)$ are important.

\section{Conclusions}
\label{sec:concl}

Three different approaches to solve the BTE beyond the RTA have
been used to study the thermal conductivity of four single-layer
TMDs: MoS$_2$, MoSe$_2$, WS$_2$ and WSe$_2$. We have shown that
$6 \times 6 \times 1$ supercells, anharmonic interactions up to
5$^\textrm{th}$-6$^\textrm{th}$ neighbors and a {\bf q}-point grid
of $60 \times 60 \times 1$ are necessary to obtain converged
values of the thermal conductivity for a sample size of 1~$\mu$m.
When available, the experimental
data agree well with our results, although the iterative
solution results in a non negligible underprediction in
the case of MoS$_2$. Also, we see that the RTA significantly 
underpredicts the $\kappa$ values for MoS$_2$ and WS$_2$.
The temperature dependence is standard,
where $\kappa$ initially increases due to a higher number of
phonons present, peaking in the 50-100~K range, and then decreasing
as Umklapp phonon-phonon scattering takes over.
Also, the spectral thermal conductivity indicates that almost
all the contribution to the room temperature $\kappa$ comes from acoustic phonons.
In addition, we found a difference of more than one order of magnitude
between the Normal and the Resistive relaxation times for all
the materials investigated, a fact that could lead to the emergence of
hydrodynamic effects similar to those already observed in graphene.
WSe$_2$, in particular, appears to be the best candidate to
observe such effects.

\section*{Acknowledgments}

We acknowledge financial support by the Ministerio de Econom\'ia, 
Industria y Competitividad (MINECO) under grants FEDER-MAT2017-90024-P, 
TEC2015-67462-C2-1-R (MINECO/FEDER) and TEC2015-67462-C2-2-R 
(MINECO/FEDER), the EU Horizon2020 research and
innovation program under grant No. GrapheneCore2 785219, and the Severo 
Ochoa Centres of Excellence Program under Grant SEV-2015-0496 and by the 
Generalitat de Catalunya under grants 2017 SGR 1506 and 2017-SGR-1018.

\section*{References}
\bibliographystyle{iopart-num}
\bibliography{./references}

\end{document}